\documentclass[twoside]{article}

\usepackage{qic,amsmath,amssymb,amsbsy,graphicx,subfigure}
\usepackage[latin1]{inputenc}

\renewcommand{\c}{\mathrm{c}}
\newcommand{\s}{\mathrm{s}}

\newcommand{\id}{\mathbb{I}}

\newcommand{\be}{\begin{equation}}
\newcommand{\ee}{\end{equation}}
\newcommand{\bea}{\begin{eqnarray}}
\newcommand{\eea}{\end{eqnarray}}

\renewcommand{\det}{{\rm Det}\,}
\newcommand{\gr}[1]{\boldsymbol{#1}}
\newcommand{\ket}[1]{|#1\rangle}

\newcommand{\sig}{\gr{\sigma}}
\newcommand{\eps}{\gr{\varepsilon}}

\newcommand{\eq}[1]{Eq.~(\ref{#1})}

\newcommand{\ie}{\emph{i.e.}~}

\begin{document}
\setlength{\textheight}{8.0truein}    

\runninghead{Correlation loss and multipartite entanglement across a black hole horizon}
            {G. Adesso and I. Fuentes-Schuller}

\normalsize\textlineskip
\thispagestyle{empty}
\setcounter{page}{657}

\copyrightheading{9}{7\&8}{2009}{0657--0665}

\vspace*{0.88truein}

\alphfootnote

\fpage{657}

\centerline{\bf
CORRELATION LOSS AND MULTIPARTITE ENTANGLEMENT}
\vspace*{0.1cm}
 \centerline{\bf ACROSS A BLACK HOLE HORIZON}
\vspace*{0.37truein}
\centerline{\footnotesize
GERARDO ADESSO}
\vspace*{0.015truein}
\centerline{\footnotesize\it School of Mathematical Sciences, University of Nottingham}
\vspace*{0.1cm} \centerline{\footnotesize\it University Park,  Nottingham NG7 2RD, UK.}
\vspace*{10pt}
\centerline{\footnotesize
IVETTE FUENTES-SCHULLER\footnote{Published before under maiden
name Fuentes-Guridi.}}
\vspace*{0.015truein}
\centerline{\footnotesize\it Institute for Theoretical Physics, Technical University of Berlin}
\vspace*{0.1cm} \centerline{\footnotesize\it
Hardenbergstr. 36, D-10623, Berlin, Germany}
\vspace*{0.035truein}
\centerline{\footnotesize\it Quantum Physics Group, STRI, School of Physics, Astronomy and Mathematics}
\vspace*{0.1cm} \centerline{\footnotesize\it
University of Hertfordshire, Hatfield, Herts AL10 9AB, United Kingdom}
\vspace*{0.225truein}
\publisher{September 16, 2008}{March 20, 2009}

\vspace*{0.21truein}

\abstracts{We investigate the Hawking effect on entangled fields. By
considering a scalar field which is in a two-mode squeezed state
from the point of view of freely falling (Kruskal) observers
crossing the horizon of a Schwarzschild black hole, we study the
degradation of quantum and classical correlations in the state from
the perspective of Schwarzschild observers confined
outside the horizon. Due to monogamy constraints on the entanglement
distribution, we show that the lost bipartite entanglement is
recovered as multipartite entanglement among modes inside and
outside the horizon. In the limit of a small-mass black hole, no
bipartite entanglement is detected outside the horizon, while the
genuine multipartite entanglement interlinking the inner and outer
regions grows infinitely.
}{}{}

\vspace*{10pt}

\vspace*{3pt}
\communicate{S Braunstein~\&~E Polzik}

\vspace*{1pt}\textlineskip    

\section{Introduction}

The physics of black holes has attracted and fascinated scientists
since almost eighty years. Recently, in the quest to understand and
properly describe such singular objects, new insights have been
drawn into and from apparently unrelated areas, such as condensed
matter physics and information theory. On the one hand, inspired by
the holographic principle, the investigation of the ``area law''
entropy scaling in  Hamiltonian lattice systems has unveiled general
connections between decay of correlations, spectral gap, and
computational complexity of ground states \cite{arealaw}. On the
other hand, the concepts of entropy, information, and entanglement
have been developed into the novel field of quantum information
science, whose applications are revolutionizing modern
communications \cite{nielsen}. It seems interesting to investigate
whether the novel tools developed in these areas may prove useful,
in turn, to achieve a better comprehension of traditional black hole
questions \cite{infoblack}.

In particular, quantum entanglement has been recognized to play an
important role in black hole thermodynamics \cite{bombelli} and in
the information loss problem in a black hole \cite{loss}, one of the
most challenging issues in theoretical physics at the moment
\cite{paradox}. Consider a snapshot of the gravitational collapse of
a massive object forming a Schwarzschild black hole. A scalar field
which is in the so-called Unruh vacuum state \cite{information} from
the point of view of freely falling observers at the horizon
(Kruskal observers) \cite{noteobservers}, contains quantum
correlations between the field modes inside and outside the horizon.
Namely, the modes described by observers living outside the
black hole (outer Schwarzschild observers) are entangled with the
modes described by virtual observers confined to the interior of the
black hole (inner Schwarzschild observers) \cite{birelli}. Such
entanglement is directly related to the entropy of the black hole
\cite{bombelli}. Schwarzschild observers outside the black hole have
no access to the modes inside the horizon, thus have incomplete
information about the quantum field, which results in the detection
of a thermal state. The so-called Hawking temperature of such a
state is inversely proportional to the black hole mass $M$
\cite{information}. For a  black hole with small, asymptotically
vanishing mass (corresponding
 e.g.~to a snapshot taken close to the final evaporation stage),
 the radiation temperature approaches infinity, resulting in
the detection of a maximally mixed state which contains no
information. This is the core of the `paradox' \cite{paradox}:
according to Hawking's  analysis, a black hole can ``swallow'' the
information about pure quantum states and eventually disappear
alongwith it.

Here we face an intriguing question: what if an {\em entangled}
state of the field is detected by infalling observers? Will the
black hole ``swallow'' also the information encoded in {\em
correlations} between the field modes?  We provide answers to these
and related questions in the following.
Let us remark that in the canonical studies of the Hawking radiation \cite{birelli}, the state of the
scalar field, from the perspective of free falling observers, is the
vacuum state.  Since the vacuum state contains no entanglement from
this perspective, it is not possible to address the question of how much
{\em quantum} information (encoded in correlations between field modes) is lost from the perspective of physical observers living outside the black hole. However, in principle, and quite realistically, the field from the Kruskal
perspective could be in any state. The most general case would
correspond to multiparty entangled states of an arbitrary number $N$
of field modes.  Unfortunately, the entanglement content of general
multipartite states cannot be exactly quantified with the existing
entanglement measures, and even its qualification is a quite formidable task.  In our analysis we take the first step beyond the conventional studies by considering
a scalar field in a
1+1 dimensional Schwarzschild black hole spacetime, which is in a
two-mode squeezed entangled state from the point of view of (freely
falling) Kruskal observers. This represents one of
the simplest nontrivial instances of $N$ entangled modes (being, in fact, $N=2$). Importantly, for the studied case
analytical tools have been developed to quantify bipartite entanglement and its redistribution into multipartite form when the field is observed from a different perspective \cite{myreview}. Therefore our analysis  generalizes the standard
studies where the tools of quantum information theory were applied to
black hole physics \cite{bombelli,loss,paradox}. The canonical studies are actually recovered when the entanglement between the two
modes vanishes from the Kruskal perspective, in which case two
independent Kruskal vacuum field modes are detected as equally
thermalized states by Schwarzschild observers.   But the most interesting contribution of our study is that considering the
two-mode squeezed state enables us to address novel questions such as
what is the effect of a black hole on the correlations (quantum and
classical) present in a multimode field. The specific choice of a two-mode
squeezed state as entangled state of the field as detected by Kruskal
observers is very natural, also in view of the ubiquity of two-mode
squeezed states in black hole physics (both in connection to its
entropy \cite{bombelli}, and to its final state \cite{ahnprd})
\cite{notetms}. Moreover, in curved spacetimes with at least two
asymptotically flat regions, particle states \cite{ball} commonly
correspond to multiparty squeezed entangled states of an arbitrary
number $N$ of field modes. Such is the case, as an example, in a
Robertson-Walker Universe. Our setting deals with the paradigmatic case $N=2$.

We show analytically that the
Kruskal correlations, both quantum and classical, in the state, are smaller
when detected by Schwarzschild observers outside the
horizon; namely, they are more intensively degraded for smaller
black hole mass $M$ and lower mode frequencies.
In the
limit $M\rightarrow 0$, quantum correlations vanish for all
frequency modes, but some classical correlations remain in the
state, although smaller than those detected by Kruskal observers.
An original insight is provided for the Hawking mechanism: we
demonstrate that the entanglement loss is precisely reciprocated by
a redistribution of the Kruskal bipartite entanglement into genuine
multipartite quantum correlations between Schwarzschild modes inside
and outside the black hole. In the limit $M\rightarrow 0$,  the
detection of an even {\em finite} bipartite entanglement by
infalling  observers, amounts to the creation of an {\em unbounded}
multipartite entanglement across the horizon. Our results thus shed
new light on the nature of the information lost in black holes,
within the quantum-informational framework of {\em monogamy} of
entanglement \cite{monogamy}.

\section{The setting}

 Let us briefly recall the Hawking effect on a scalar field. Each Unruh vacuum mode of
frequency $\alpha$, denoted by $\ket{0}^{\cal K}_{\alpha}$ in the
Kruskal frame, corresponds to a two-mode squeezed state in the
Schwarzschild frame \cite{noteobservers,information,birelli} (see
also Ref.~\cite{ahnprd}),
\begin{eqnarray}\left| 0\right\rangle^{\mathcal{K}}_{\alpha} &=&
\frac{1}{\cosh r} \sum_{k=0}^{\infty }\tanh ^{k}r\,\left|
k\right\rangle _{\alpha}^{in}\left| k\right\rangle
_{\alpha}^{out}\nonumber\\&\equiv&U_{\alpha^{in},\alpha^{out}}(r)\left|
0\right\rangle _{\alpha}^{in}\left| 0\right\rangle
_{\alpha}^{out}\,.\label{bogo}\end{eqnarray} Here $\left|
k\right\rangle _{\alpha}^{in}$ and $\left| k\right\rangle
_{\alpha}^{out}$ are the Fock states inside and outside the black
hole, $U_{i,j}(r)=\exp \frac{r}{2} (\hat {a}_i^\dag \hat {a}_j^\dag
-\hat {a}_i \hat {a}_j )$ is the two-mode squeezing operator with
$r$ the (real) squeezing parameter, and $a_{i,j}$,
$a_{i,j}^{\dagger}$, are creation and annihilation operators for
modes $i,j$.
 The squeezing  $r$ is related to
the frequency $\alpha$ and the mass $M$ of the black hole (in
natural units),
\begin{equation}\label{accparam} \cosh r
=\left(1-e^{-2M\pi\alpha}\right)^{-1/2}\,.
\end{equation}

The two-mode squeezed state of \eq{bogo} is a {\em Gaussian state},
\ie a continuous variable state with Gaussian characteristic
function \cite{myreview}. Up to local unitary operations, all the
information about a Gaussian state is contained in the symmetric
covariance matrix (CM) whose entries are
$\sigma_{ij}=1/2\langle\{\hat{X}_i,\hat{X}_j\}\rangle
-\langle\hat{X}_i\rangle\langle\hat{X}_j\rangle$ where
$\hat{X}=\{\hat x_1,\hat p_1,\ldots,\hat x_N,\hat p_N\}$ is the
vector of the field quadrature operators. In terms of CMs, \eq{bogo}
reads $\sig^{(0){\cal K}}_{\alpha} =
\Gamma_{\alpha^{in},\alpha^{out}}(r) (\sig^{(0){in}}_{\alpha} \oplus
\sig^{(0){out}}_{\alpha})\Gamma^T_{\alpha^{in},\alpha^{out}}(r)$,
where $\sig^{(0){in}}_{\alpha}=\sig^{(0){out}}_{\alpha}=\id_2$ and
$\Gamma_{i,j}(r)$ is the symplectic transformation associated to the
two-mode squeezing operator $U_{i,j}(r)$,
\begin{equation}\label{tms}
\Gamma_{i,j}(r)=\left(\begin{array}{cc}
\cosh(r) \id_2&\sinh(r) Z_2\\
\sinh(r) Z_2&\cosh(r) \id_2
\end{array}\right)\! , \hbox{ with $Z_2={{1\ \ \ 0}\choose {0 \ -1}}$}.\nonumber
\end{equation}

Suppose now to have, from the perspective of Kruskal observers, an
entangled two-mode squeezed state, which is described by the CM
$\sig^{\cal
K}_{\lambda\nu}(\xi)=\Gamma_{\lambda\nu}(\xi)\Gamma^T_{\lambda\nu}(\xi)$,
$\lambda$ and $\nu$ being the frequencies of the modes.
 The entropy of entanglement \cite{vneunote} between the modes $\lambda$ and
 $\nu$ is $S_\xi=f(\cosh 2\xi)$, where
\begin{equation}\label{f}
f(x) \equiv \frac{x+1}{2} \log\left[\frac{x+1}{2}\right]-
\left(\frac{x-1}{2}\right)\log\left[\frac{x-1}{2}\right].
\end{equation}
The two modes also share classical correlations, equal in content to
the entanglement $S_\xi$. This is because the mutual information
\cite{berry}, quantifying the total (quantum plus classical) amount
of correlations in the state, is equal to $I_\xi=2S_\xi$. In the
limit of infinite squeezing ($\xi\to \infty )$, the state approaches
the ideal (unnormalizable) EPR state \cite{epr}, which contains an
infinite amount of entanglement and mean energy.

We now wish to describe the state from the perspective of
Schwarzschild observers which have only access to the field outside
the black hole horizon. Each Kruskal mode is thus transformed
according to the Bogoliubov transformation which acts on the vacuum
as in \eq{bogo} \cite{information,birelli}(see also \cite{ahn1}).
The global state, involving modes inside and outside the black hole,
will be thus a four-mode pure Gaussian state with CM
$\sig_{\lambda\nu}^{in,out} = O O^T$, where
$O=\Gamma_{\nu^{in},\nu^{out}}(n)
\Gamma_{\lambda^{in},\lambda^{out}}(l)\Gamma_{\lambda\nu}(\xi)$, and
the parameters $l$ and $n$ are related to the black hole mass $M$
and to the mode frequencies $\lambda, \nu$ respectively, by
\eq{accparam}. Explicitly,
\begin{equation}\label{sig4}
\sig_{\lambda\nu}^{in,out}=\left(%
 \begin{array}{cccc}
  \sig_{\lambda^{in}} & \eps_{\lambda^{in}\lambda^{out}} &  \eps_{\lambda^{in} \nu^{out}}  & \eps_{\lambda^{in} \nu^{in}}\\[0.2cm]
  \eps^T_{\lambda^{in}\lambda^{out}} & \sig_{\lambda^{out}} & \eps_{\lambda^{out} \nu^{out}} & \eps_{\lambda^{out} \nu^{in}}\\[0.2cm]
  \eps^T_{\lambda^{in} \nu^{out}}  & \eps^T_{\lambda^{out} \nu^{out}} & \sig_{\nu^{out}} & \eps_{\nu^{out} \nu^{in}} \\[0.2cm]
  \eps^T_{\lambda^{in} \nu^{in}} & \eps^T_{\lambda^{out} \nu^{in}} & \eps^T_{\nu^{out} \nu^{in}} &
  \sig_{\nu^{in}}
\end{array}%
\right)\,,
\end{equation}where (setting $\c=\cosh$ and $\s=\sinh$):
\begin{eqnarray*}
 \sig_{X^{in}} &\!\!=\!\!& [\c^2(x) + \c(2 \xi) \s^2(x)] \id_2\,, \\
 \sig_{X^{out}} &\!\!=\!\!& [\c^2(x)\c(2\xi)+\s^2(x)] \id_2\,, \\
\eps_{X^{in} X^{out}}=\eps_{X^{out} X^{in}} &\!\!=\!\!& [\c^2(\xi)
\s (2 x)]
Z_2\,,\\
\eps_{X^{in} Y^{out}}=\eps_{Y^{out} X^{in}} &\!\!=\!\!& [\c (y) \s
(2 \xi) \s
(x)] \id_2 \,, \\
\eps_{X^{in} Y^{in}}&\!\!=\!\!&[\s (2\xi) \s(x) \s (y)] Z_2\,,\\
\eps_{X^{out} Y^{out}} &\!\!=\!\!& [\c (x) \c (y) \s (2 \xi)] Z_2\,.
\end{eqnarray*}
Here $X,Y=\{\lambda,\nu\}$ and accordingly $x,y=\{l,n\}$.

\section{Bipartite entanglement and correlations between the outer modes}

 The observers moving outside the black hole have access only to
modes ${\lambda}^{out}$ and ${\nu}^{out}$. Therefore, after tracing
over the states in the region inside the black hole, the resulting
state is mixed. We can exactly quantify entanglement in different
partitions of (mixed) Gaussian states by using a measure known as
the {\em contangle} $\tau$ \cite{notecontangle}.  The contangle is
equivalent to the Gaussian entanglement of formation \cite{geof},
which quantifies the cost of creating a mixed Gaussian state out of
an ensemble of pure entangled Gaussian states. In the pure-state
case, the entanglement of formation reduces to the entropy of
entanglement \cite{vneunote}. The entanglement between the modes
$\lambda^{out}$ and $\nu^{out}$ from the perspective of observers
outside the black hole is given by $\tau_{\lambda|\nu} \equiv
\tau(\sig^{out}_{\lambda\nu})=g[m^2_{\lambda|\nu}]$
\cite{notecontangle}, where
\begin{eqnarray}
m_{\lambda|\nu} =\frac{2 \c(2l) \c(2n) \c^2(\xi) + 3\c(2\xi) -
      4\s(l) \s(n) \s(2\xi) -
      1}{2 \left[(\c(2l) + \c(2n)) \c^2(\xi) - 2\s^2(\xi) +
          2 \s(l) \s(n) \s(2 \xi)\right]},\nonumber
\end{eqnarray}
in case $\tanh (\xi)> \sinh (l) \sinh (n)$. Otherwise, the
entanglement vanishes.

\begin{figure}[tb!]
\centering{\includegraphics[width=10cm]{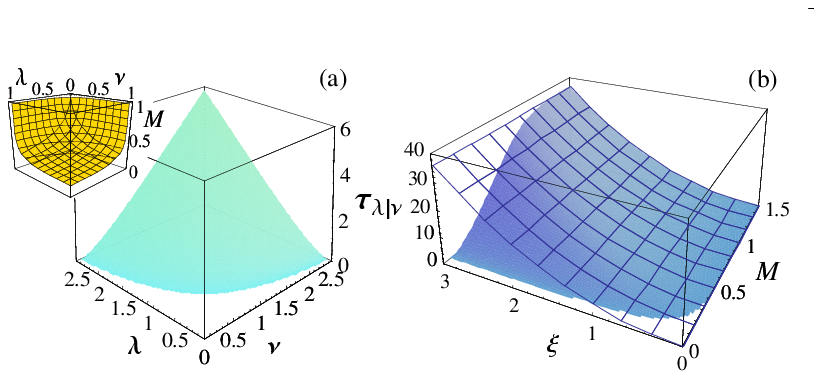}%
\fcaption{\label{fig:ent} Entanglement between two Schwarzschild
modes detected outside a black hole (only nonzero values are
plotted). (a) Contangle $\tau_{\lambda|\nu}$ as a function of the
mode frequencies, for a black hole of mass $M=1/(2\pi)$ and infinite
entanglement ($\xi \rightarrow \infty$) in the Kruskal frame. Inset:
trade-off between frequencies and mass for $\xi \rightarrow \infty$;
entanglement is nonzero only inside the convex meshed surface. (b)
Contangle $\tau_{\lambda|\nu}$ (shaded surface) as a function of
squeezing $\xi$ and  mass $M$, for $\lambda=1$ and $\nu=2$, compared
to the contangle $4\xi^2$ (wireframe surface) detected by Kruskal
observers.}}
\end{figure}

The bipartite entanglement depends on the degree of squeezing $\xi$
(which determines the entanglement described by Kruskal observers),
on the frequencies $\lambda$ and $\nu$, and on the mass $M$ of the
black hole. As shown in Fig.~\ref{fig:ent}, entanglement increases
with all these parameters. However, as function of the effective
squeezing parameters $l$ and $n$ entanglement is strictly
decreasing. We observe that the degradation on entanglement with
decreasing mass is stronger if the entanglement $S_\xi$ in the
Kruskal frame is larger. At fixed $\lambda$ and $\nu$, it is
interesting to see that entanglement vanishes for a finite value of
$M$.
Conversely, note that for a fixed black hole mass $M$, the
entanglement of low frequency modes vanishes or, better said,
becomes physically unaccessible to the observers outside the black
hole. Even in the ideal case, where the Kruskal state is an
infinitely entangled EPR state \cite{epr} (corresponding to $\xi
\rightarrow \infty$), the entanglement between modes $\lambda$ and
$\nu$ {\em completely} vanishes if $e^{2\pi\lambda M}+e^{2\pi\nu
M}-e^{2\pi M(\lambda+\nu)}\geq 0 $ [see inset of
Fig.~\ref{fig:ent}(a)]. Only modes with high frequency remain
entangled. Moreover, the number of entangled modes becomes smaller
for decreasing black hole mass and, as expected, the pairwise
entanglement between two modes with given frequencies becomes
weaker. In the limit $M\rightarrow 0$ the bipartite entanglement
exactly vanishes between {\em all} frequency modes. The quantum
correlations detected by Schwarzschild observers outside the black
hole are irremediably lost, as we have just
 demonstrated in a full analytical fashion.

\begin{figure}[tb!]
\centering{\includegraphics[width=8cm]{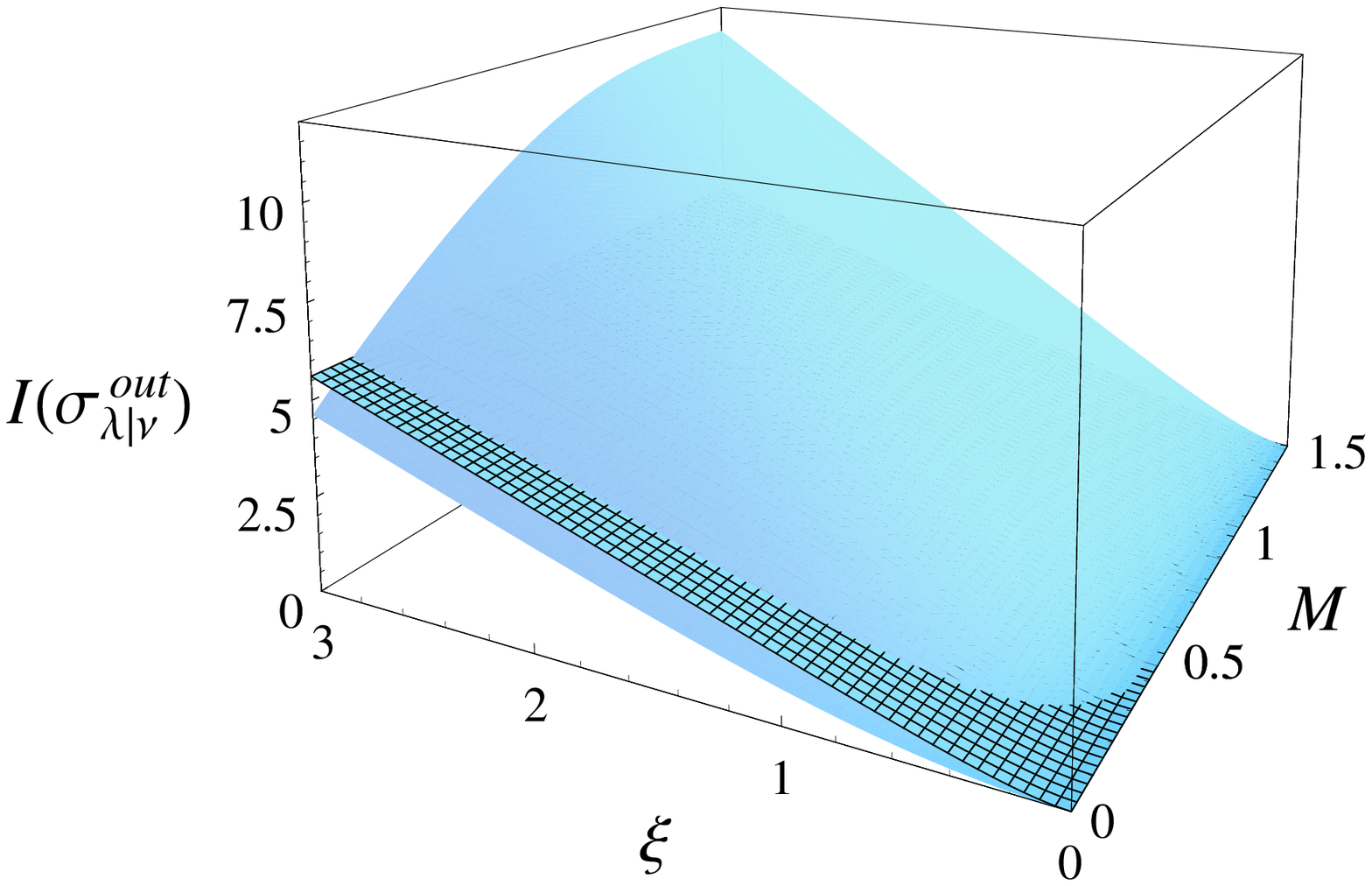}%
\fcaption{\label{fig:class} Total correlations
$I(\sig_{\lambda\nu}^{out})$ (shaded surface) between modes with
frequency $\lambda=1$ and $\nu=2$ detected outside a black hole,
versus mass $M$ and squeezing $\xi$. The classical correlations
$S_\xi$ detected by Kruskal observers are displayed as well (meshed
surface).}}
\end{figure}

It is also possible to calculate the total correlations between the
modes outside the black hole using the mutual information
\cite{berry} which in terms of the CM reads \cite{serafini}
$$I(\sig_{\lambda\nu}^{out}) =
f(\sqrt{\det\sig_{\lambda^{out}}})+f(\sqrt{\det\sig_{\nu^{out}}}) -
f(\eta_{\lambda\nu}^+)- f(\eta_{\lambda\nu}^-)\,,$$ where $f(x)$ is
defined in \eq{f}, and the symplectic eigenvalues are given by
\cite{extremal} $$2({\eta_{\lambda\nu}^{\pm}})^2 =
\Delta(\sig_{\lambda\nu}^{out})\pm
\sqrt{\Delta(\sig_{\lambda\nu}^{out})^2-4\det(\sig_{\lambda\nu}^{out})}\,,$$
with
\begin{eqnarray*}
\det(\sig_{\lambda\nu}^{out})&\!=\!& [\c^2(n) + \c(2 \xi) \s^2(n)]
\c^2(l) + \s^2(l) [\c(2 \xi) \c^2(n) + \s^2(n)]^2\,, \\ \Delta(\sig_{\lambda\nu}^{out}) &\!=\!& \c^2(2 \xi) \c^4(l) +
  2 [\c (2 \xi) \s^2(l) - \c^2(n) \s^2(2 \xi)] \c^2(l) + \s^4(l) + [\c (2 \xi) \c^2(n) + \s
^2(n)]^2\,.\end{eqnarray*} The behavior of the mutual information is qualitatively
similar to that of the bipartite entanglement. The most remarkable
and original feature is that, in the regime where low frequency
modes are disentangled (corresponding to a black hole with small
mass, $M \rightarrow 0$), classical correlations are degraded as
well. Namely (see Fig.~\ref{fig:class}), the mutual information
$I(\sig_{\lambda\nu}^{out})$ may fall below the threshold
$S_\xi=f(\cosh 2\xi)$, which corresponds to the classical
correlations (and, equivalently, the entanglement) detected by
Kruskal observers. Previous analyses reported classical correlations
invariant under the Hawking effect, by considering either a
maximally entangled two-qubit Kruskal state \cite{ahn1}, or the
correlations between modes measured by an observer falling into the
black hole and the other escaping the fall \cite{rindler,prd}. We
find instead that, for a two-mode squeezed Kruskal state,
 classical correlations
are strictly invariant only in the limit of infinite entanglement,
$\xi \rightarrow \infty$, in which case for $M \rightarrow 0$, the
total correlations $I(\sig_{\lambda\nu}^{out}) \rightarrow S_\xi$.
For any finite degree of squeezing $\xi$ (which only then is
physically meaningful), we have thus shown that Hawking radiation
affects mostly quantum, but also classical correlations between
modes outside a black hole. We observe that, nevertheless, the loss
of classical correlations never exceeds one ebit.

\section{Genuine multipartite entanglement shared by inner and outer modes}

We are interested in understanding what happened to the lost quantum
correlations. Are they really lost?
Since the two entangled Kruskal modes are transformed into a
four-partite system from the Schwarzschild perspective (with two
modes, one inside and another outside the black hole, for each
frequency $\lambda$ and $\nu$) we move to analyze the information
stored in form of multipartite correlations in the state.
Multipartite entanglement is a complex topic which can be addressed
in very few cases, which fortunately include, thanks to some recent
advances \cite{contangle,hiroshima,unlim}, Gaussian states. The
measure of multipartite entanglement we use is called the
\emph{residual contangle} and stems from the observation that
entanglement cannot be freely shared. There are `monogamy' relations
which impose trade-offs on the distribution of bipartite
entanglement between the various mode partitions of the system
\cite{monogamy}, constraining the residual multipartite entanglement
 not stored in couplewise form. The minimum residual
entanglement in the state $\sig_{\lambda\nu}^{in,out}$, derived from
the entanglement monogamy inequality defined in
\cite{contangle,hiroshima,unlim}, is given by the contangle between
the mode $\lambda^{in}$ and the rest, minus the contangle between
$\lambda^{in}$ and $\lambda^{out}$, i.e.,
$\tau^{res}(\sig_{\lambda\nu}^{in,out})=\tau_{\lambda^{in}|(\lambda^{out}\nu^{in}\nu^{out})}-\tau_{\lambda^{in}|\lambda^{out}}$.
Here we labeled the modes such that $l \le n$ without loss of
generality, and used the fact that there are no quantum correlations
between modes inside and outside  of different frequency, nor
between $\lambda^{in}$ and $\nu^{in}$. Recalling that
$\tau_{\lambda^{in}|\lambda^{out}}=4l^2$ and
$\tau_{\nu^{in}|\nu^{out}}=4 n^2$, the residual contangle reads
\begin{equation} \label{taures} \tau^{res} = {\rm
arcsinh}^2\!\left\{ {\sqrt {[\cosh ^2l+\cosh (2\xi)\sinh ^2l]^2-1} }
\right\}-4l^2.
\end{equation}
 The derivation of this
result is a straightforward generalization of the findings of Ref.
\cite{unlim}, where the multipartite entanglement of the state in
\eq{sig4} has been studied, for $l=n$, in a quantum-optical setting.
All technical details of these calculations are given in Refs.
\cite{unlim,prd}.

This residual entanglement contains both tripartite and genuine
four-partite contributions. The portion of tripartite correlations
in the state, with respect to the above mode decomposition, only
involves the mode $\lambda^{in}$ and the two modes outside the black
hole (the other contributions being zero). We calculate an upper
bound to such tripartite entanglement between modes $\lambda^{in}$,
$\lambda^{out}$ and $\nu^{out}$ using the techniques of Refs.
\cite{unlim,prd} and obtain, $$\begin{array}{c}\tau^{tri} \le
\min\big\{g\big[\big(\frac{\mathrm{sech}^2(n) \tanh ^2(\xi) +
                      1}{\mathrm{sech}^2(n) \tanh ^2(\xi) -
                      1}\big)^2\big] - g[m_{\lambda | \nu }^2],\
     g\big[\big(\frac{\mathrm{sech}^2(n) \tanh ^2(\xi) - \cosh (2 l)}{
     \mathrm{sech}^2(n) \tanh ^2(\xi) - 1}\big)^2\big] -4l^2\big\}\,.\end{array}$$
   The tripartite correlations vanish asymptotically for $l, n \gg 0$, which
   means that a portion of the information in the Kruskal state is
     redistributed into correlations with the remaining mode inside,
     $\nu^{in}$, which are neither of bipartite nor of tripartite nature.
     Remarkably,  in the regime of
     large $l, n$ (low frequencies and/or small black hole mass), we conclude that the
     residual entanglement given by \eq{taures} quantifies exclusively the
     genuine four-partite quantum correlations among all four modes.

\begin{figure}[tb!]
\centering{\includegraphics[width=8cm]{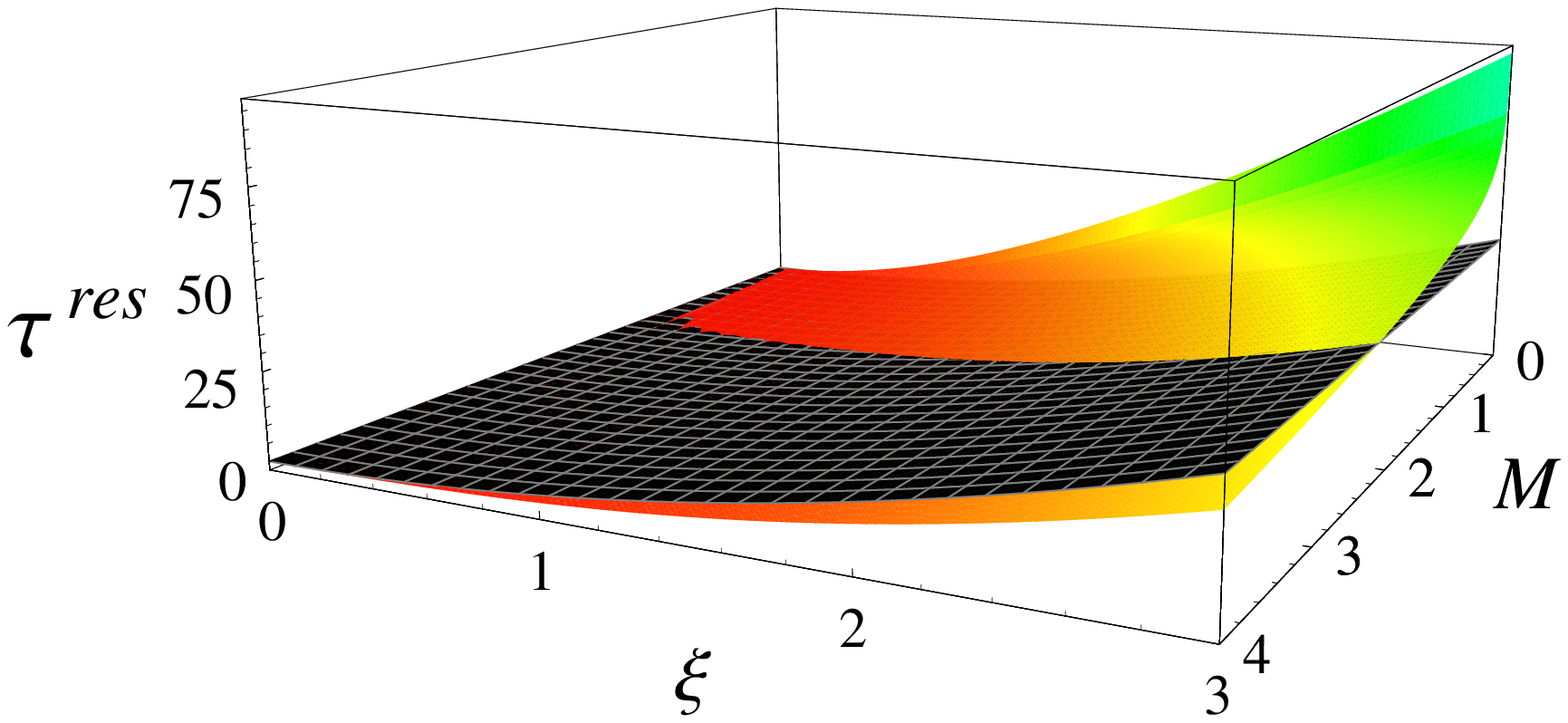}%
\fcaption{\label{fig:multi} Residual multipartite entanglement among
the four (in and out) modes, \eq{taures} (shaded surface), versus
squeezing $\xi$ and black hole mass $M$, at fixed $\lambda=1/(8
\pi)$. The bipartite entanglement $4\xi^2$ of the Kruskal state is
displayed as well (meshed surface).}}
\end{figure}

     We plot the residual entanglement in Fig.~\ref{fig:multi} where
     we see that it grows with the squeezing $\xi$. For fixed $\xi$, it
{\em increases} with
     decreasing black hole mass $M$. Interestingly, the four-partite entanglement can easily surpass the
     bipartite entanglement $4\xi^2$ in the Kruskal state, and diverges in the
     limit $M \rightarrow 0$. Therefore, not only do
     we find that all the bipartite correlations in the Kruskal state are entirely
     redistributed into four-partite correlations
     among the Schwarzschild modes, but also that an
     unlimited amount of four-mode correlations is created by the presence of a black
     hole with infinitesimal mass. In this limit, the entanglement between modes
of the same frequency across the horizon also diverges: this partly
explains the generation of infinite genuine multipartite
entanglement among all the four modes (the entanglement is said to
distribute in a monogamous but  `promiscuous' way \cite{unlim}) and
why the bipartite entanglement between the outside modes is degraded
and, with decreasing mass, destroyed. The infinite energy,
associated with such an infinite multipartite
 entanglement interlinking the inside
and outside regions of the black hole horizon, is related to the
infinite temperature at which the black hole radiates, according to
the Hawking effect \cite{information}, in the small-mass limit
\cite{notebreak}.

\section{Conclusions}

In this paper, we have  characterized the nature of information loss
in the case of entangled bosonic fields in the presence of a
Schwarzschild black hole horizon. A novel, interesting signature of
the Hawking mechanism \cite{information} is derived: the detection
of an even {\em finite} bipartite entanglement by infalling
observers, amounts to the creation of an {\em unbounded}
multipartite entanglement across the horizon. The limiting situation
of a black hole with vanishing mass (a snapshot of its evaporation),
commonly regarded as the origin of the information loss `paradox'
\cite{paradox}, is precisely identified by a divergence of such
multipartite entanglement \cite{notebreak}. Already at finite black
hole masses, observers outside the horizon may witness (depending on
the mode frequencies) absolutely no entanglement in the field, and a
small degradation of classical correlations as well. Connection of
these insights to concepts of black hole thermodynamics awaits
further investigation.

\section*{Acknowledgments}
\noindent We thank F. Illuminati and M. Ericsson for discussions. I. F-S was supported by the Alexander von Humboldt Foundation and would like to thank Tobias Brandes for his hospitality.


\begin{thebibliography}{99}
\bibitem{noteobservers} Kruskal (Schwarzschild) observers are defined by
the vector field $\partial u$ ($\partial t$), where $u$ ($t$) is the
Kruskal (Schwarzschild) temporal coordinate
\cite{birelli,information}. A Kruskal (Schwarzschild) state/mode
refers to the state/mode described by the respective observers.
\bibitem{notetms} In this respect we notice that the degradation of correlations in a different type of two-mode entangled Kruskal state, i.e. a two-qubit-like
Bell state $\ket{00}+\ket{11}$, as detected by Schwarzschild
observers, have been recently studied \cite{ahn1}. However, in that case it was not possible to frame the effect of black hole within the theory of entanglement redistribution, nor a degradation of classical correlations was observable.
\bibitem{vneunote} The entropy of entanglement of a
bipartite state $\ket{\psi}_{AB}$ is defined as
$S(\varrho_{A})=S(\varrho_{B})=-{\rm
tr}[\varrho_{A}\log(\varrho_{A})]$, where $\varrho_{A}$
($\varrho_{B}$) is the reduced density matrix of subsystem $A$
($B$).
\bibitem{notecontangle}
The contangle is  defined as $\tau (\sig_{i\vert j} )\equiv
g[m_{i\vert j}^2 ]$ where $g[x]={\rm arcsinh}^2[\sqrt {x-1} ]$ and
$\sig_{i\vert j} $ is the CM of a (mixed) bipartite Gaussian state
where subsystem $i$ comprises one mode only \cite{contangle}. Here
$m_{i\vert j}\!\equiv\!m(\sig _{i\vert j}^{opt})\!=\!(\det
\sig_i^{opt})^{1/2}$, where $\sig_{i}^{opt}$ is the reduced CM of
subsystem $i$. The CM $\sig_{i\vert j}^{opt} $ denotes the pure
bipartite Gaussian state which minimizes $m(\sig_{i\vert j}^p )$
among all pure-state CMs $\sig_{i\vert j}^p \le \sig_{i\vert j}$.
Separability amounts to $m_{i\vert j}=1$.
\end{thebibliography}
\end{document}